\documentclass[10pt,twocolumn,preprintnumbers,amsmath,amssymb,nofootinbib
,superscriptaddress]{revtex4-2}
\usepackage{graphicx,longtable,mathrsfs,color,array}
\usepackage{hyperref}
\usepackage[usenames,dvipsnames]{xcolor} 
\usepackage{amssymb,amsmath,mathtools,mathrsfs,slashed} 
\usepackage{epsfig,subfigure,placeins,float} 
\usepackage{booktabs,longtable,ctable,multirow} 
\usepackage{exscale,relsize} 
\usepackage[normalem]{ulem} 
\usepackage{enumerate}
\usepackage{enumitem}
\usepackage{comment}
\usepackage{color}
\usepackage{braket}
\allowdisplaybreaks[1]

\newcommand{\be}{\begin{equation}}
\newcommand{\ee}{\end{equation}}
\newcommand{\bea}{\begin{eqnarray}}
\newcommand{\eea}{\end{eqnarray}}

\newcommand{\pin}{{\mbox{\tiny PI}}}
\newcommand{\pv}{{\mbox{\tiny PV}}}
 \begin{document}
\title{Gravitational-Wave Propagation Through the Axiverse}

\author{Leah Jenks}
\email{ljenks3@jh.edu}
\affiliation{William H. Miller III Department of Physics \& Astronomy,\\ Johns Hopkins University, 3400 N. Charles St., Baltimore, MD 21218, USA}

\author{Marc Kamionkowski}
\email{kamion@jhu.edu}
\affiliation{William H. Miller III Department of Physics \& Astronomy,\\ Johns Hopkins University, 3400 N. Charles St., Baltimore, MD 21218, USA}

\begin{abstract}

We study the effects of oscillating, ultralight scalar and pseudoscalar fields on the propagation of gravitational waves (GWs). We consider two potential couplings of the (pseudo)scalars to gravity; a parity-even Gauss-Bonnet coupling, and parity-odd Chern-Simons coupling. We find several effects at both the population and individual GW event level, characterized by oscillatory features controlled by the (pseudo)scalar mass. In the parity-even case, this feature can be seen in the observed GW redshift and speed distributions, as well as in the dispersion relation and phase of individual events. We use the observation of the GW170817 multimessenger binary neutron star event to place constraints on the parity-even scalar-graviton coupling. In the parity-odd case, the effects are birefringent, but we find an overall washout of polarization at the population level. Oscillatory features can be seen in the observed GW amplitude and inclination distributions. Finally, we find that continuous, monochromatic GW sources are a promising target to observe these effects. The presence of a (pseudo)scalar field imprints a modulation of the GW waveform in the time domain, which can potentially be observed with space-based detectors such as LISA. 

\end{abstract}

\date{\today}
\maketitle

\section{Introduction}
\label{sec:intro}

Ultralight scalar and pseudoscalar fields are ubiquitous in particle physics and cosmology. These fields can arise in a variety of contexts, including as dark matter and dark energy candidates, and from string theory, which predicts an `axiverse' of many such fields \cite{Svrcek:2006yi, Masso:1995tw, Masso:2002ip, Conlon:2006tq,Arvanitaki:2009fg,Preskill:1982cy, Abbott:1982af, Dine:1982ah, Sikivie:1983ip, Ipser:1983mw, Hu:2000ke, Hui:2021tkt,Frieman:1995pm,Carroll:1998zi}. In each of these cases, it is likely that the couplings to the Standard Model (SM) are suppressed, making them difficult to detect with traditional particle physics or direct detection experiments. In this case, the dominant interactions that these fields experience are gravitational. However, if these additional new fields come from string theory or some other ultraviolet (UV) theory, they can inherit non-minimal couplings to gravity, making the gravitational sector a promising landscape to search for new physics. For this reason, there has been significant effort, with the advent of gravitational-wave (GW) astrophysics, dedicated to studying how GWs can be used as a tool to probe ultralight fields, by considering effects such as superradiance, the effects of black holes merging within dense dark matter environments, GW distortions via gravitational lensing \cite{Bertone:2019irm}, and heterodyning frequency modulation \cite{Blas:2024duy}. 

There are a variety of ways that (pseudo)scalar fields, $\varphi$, can couple to gravity in a non-minimal fashion. One particularly well-studied class are couplings of (pseudo)scalar fields to quadratic curvature invariants. These interactions arise naturally in the treatment of gravity as an effective field theory, with various operators arising in particle physics and string theory contexts. Two well-motivated examples are the coupling of a scalar to the parity-even Gauss-Bonnet invariant, $\varphi{\cal G}$, which can arise from heterotic string theory \cite{Zwiebach:1985uq, Gross:1986mw,Cano:2021rey}, and the coupling of a pseudoscalar axion to the parity-odd Pontryagin invariant, $\varphi R \tilde{R}$, which arises from anomaly cancellation in string theory and from the gravitational chiral anomaly in particle physics \cite{Green:1987mn, Lue:1998mq, jackiw, Alexander:2009tp,ALVAREZGAUME1984269,Polchinski:1998rr, PhysRevLett.96.081301, Alexander:2004xd, Alexander:2021ssr}. There is much work studying the effects of these two operators in the gravitational sector on gravitational waves and black holes when the associated ultralight field is a massless, slowly varying background field, in the context of scalar-Gauss-Bonnet gravity (see the review \cite{Fernandes:2022zrq} and references therin) and dynamical Chern-Simons gravity \cite{Jackiw:2003pm, Alexander:2009tp}. However, there is a smaller body of work exploring the effects of these couplings when the associated field is massive \cite{Alexander:2025olg,Ema:2021fdz, Dunsky:2025pvd,Chung:2026zon,Xie:2024xex,Figliolia:2025dtw}.\footnote{There also exists a body of literature studying the effects of other classes of nonminimal couplings in the context of gravitational waves and dark matter, e.g. \cite{Cai:2020ovp,Zhang:2023fhs,Cai:2023ykr,Chen:2024pyr,Zhang:2025kze}.} Given that these couplings are expected to naturally arise from string theory and a variety of other high-energy processes, their effects on GWs must be taken into account both when modeling GW waveforms and when looking for signatures of new ultralight degrees of freedom in the gravitational sector.  

In this work, we study how GW signals are modified as they propagate through a coherently oscillating ultralight field, considering both the parity-even Gauss-Bonnet coupling and the parity-odd Chern-Simons coupling. In both cases, we compute the corrections to the GW waveform induced as the GWs propagate. Our findings are applicable for GWs arising from any source, but we choose to focus on GWs from compact binaries in the late-time universe.  For the parity-even coupling, there are corrections to both the amplitude and phase (and therefore the speed) of the GWs which lead to a modulation of various observables at a frequency controlled by $m_\varphi$. At the population level, this appears as oscillations in the redshift distribution of binary black holes, as well as in the GW speed distribution from binary neutron stars. From the observation of the multi-messenger binary neutron star event, GW170817, we place novel constraints on the Gauss-Bonnet scalar-graviton coupling parameter. In the parity odd scenario, we find that there is again an amplitude modulation with oscillatory redshift dependence. This modulation can be discerned from the redshift distribution of events in polarization components, or in the observed inclination distribution. Furthermore, there is an overall washout of polarization, compared to the net circular polarization that occurs in the slowly-varying  axion background case. In both cases, we estimate the number of events required to realistically observe both the parity-even and parity-odd effects, and find that such observations are within the range of next-generation GW detectors such as Cosmic Explorer and Einstein Telescope \cite{ET:2019dnz,Reitze:2019iox}. Finally, we show that quasi-monochromatic, continuous wave sources which will be observed by space-based GW detectors such as the Laser Interferomoter Space Antenna (LISA) \cite{LISA:2017pwj} can also probe these ultralight (pseudo)scalar effects.

The structure of the paper is as follows: In Section~\ref{sec:summary}, we provide a high-level executive summary of our results. In Section~\ref{sec:even}, we show how gravitational wave propagation is altered by ultralight scalars with a parity-even Gauss-Bonnet coupling to gravity, place constraints on the coupling parameter in terms of the scalar mass, and show the characteristic modulation of the redshift and speed distributions. We estimate the necessary number of events to probe these effects. In Section ~\ref{sec:odd} we discuss the parity-odd Chern-Simons coupling, finding a modulation of the observed inclination distribution and a polarization-dependent modulation of the redshift distribution.  Then, in Section~\ref{sec:monoGW}, we show how the induced waveform modulation can be observed with quasi-monochromatic, continuous wave sources which will be seen in LISA. Finally, in Section~\ref{sec:discussion} we conclude with a discussion. Throughout the paper we use natural units such that $c=\hbar = 1$ unless otherwise specified, and we employ a mostly plus metric signature.

\section{Executive Summary}
\label{sec:summary}

In this work we study the effects of an oscillating ultralight (pseudo)scalar field on the propagation of gravitational waves. We consider a parity-even and parity-odd coupling scenario for the (pseudo)scalar to gravity, point out several physical phenomena that arise in gravitational wave observables, and discuss the optimal strategies to search for such effects. The phenomena we discuss here are distinct from those in existing work that study how (pseudo)scalars that are slowly varying
cosmological timescales affect GW propagation.\footnote{This is analogous to cosmic birefringence in the cosmic microwave background (CMB), which similarly has distinct phenomena if the background axion field is slowly varying or coherently oscillating, as discussed in \cite{Fedderke:2019ajk}. } In that situation, the modification to the GW waveform is \cite{Jenks:2023pmk}
\be 
h_{\rm R, L}^{\rm slow} = \bar{h}_{\rm R,L}\exp\left[\delta_{\rm A}^{\rm slow} f_{\rm A}(z)\right] \exp\left[i \delta_{\rm P}^{\rm slow} f_{\rm P}(z) \right],
\ee 
where $\bar{h}_{\rm R,L}^{\rm slow}$ are the right- and left-handed GW polarization modes in vacuum in the absence of an intervening field, $\delta_{\rm A}^{\rm slow}$, $\delta_{\rm P}^{\rm slow}$ are prefactors for the amplitude, and phase, respectively, which can depend on parameters such as the GW frequency, $f$, and (pseudo)scalar derivative, $\dot{\varphi}$. The functions $f_{\rm A}(z)$ and $f_{\rm P}(z)$ are monotonic functions of the redshift, $z$. For the parity-even coupling, this is known to lead to an amplitude correction, as well as a phase correction \cite{Satoh:2007gn,Daniel:2024lev}. In the parity-odd case, there is amplitude birefringence, such that either the right- or left-handed GW will get amplified and the other attenuated along the path of propagation \cite{Alexander:2007kv, Alexander:2017jmt}. 

When the (pseudo)scalar is rapidly oscillating compared to the cosmological expansion, we find a different effect. In this case, the amplitude and phase corrections to the GW will be modified to have \textit{oscillatory redshift dependence} as 
\be 
h_{\rm R, L}^{\rm osc}= \bar{h}_{\rm R,L}\exp\left\{\delta_{\rm A}^{\rm osc} \sin [f_{\rm A}(z)]\right\} \exp\left\{i \delta_{\rm P}^{\rm osc}\sin[f_{\rm P}(z)] \right\},
\ee 
where 
\be 
\sin[f_{\rm A}(z)], \, \sin[ f_{\rm P}(z)] \propto \sin m_\varphi t, 
\ee 
with $m_\varphi$ the (pseudo)scalar mass and $\delta_{\rm A}^{\rm osc}, \delta_{\rm P}^{\rm osc}$ the amplitude and phase prefactors in the oscillating field scenario. This small change in the waveform modification leads to dramatically different effects on GW observables than in the slowly varying scenario. We summarize these effects below: 

\begin{enumerate}
    \item The GW amplitudes and/or phases from binary mergers get enhanced or attenuated based on propagation distance, making it difficult to precisely characterize effects with a single event. 
    \item At the population level, the oscillatory effects will imprint themselves on the redshift distribution of various observables, including the GW amplitude, speed, and inclination. For masses, $10^{-31} \lesssim m_\varphi/{\rm eV }\lesssim 10^{-29}$, the oscillations in the redshift distribution can be observed directly. For masses $ m_\varphi / {\rm eV} \gtrsim 10^{-28}$ the oscillations are too rapid to be resolved in redshift space. However, there will be a statstically significant scatter in the redshift distribution to indicate the presence of an oscillating ultralight field. Then, analysis of individual waveforms can be used to determine the (pseudo)scalar mass and coupling.
    \item The parity-even scenario is highly constrained by the binary neutron star event, GW170817. We present novel constraints on the parity-even scalar-graviton coupling for the mass range $10^{-31} \lesssim m_\varphi/{\rm eV} \lesssim 10^{-4}$, depending on the density of the scalar field.
    \item There is an overall washout of polarization in the parity-odd case compared to the slowly varying scenario, in which we expect there to be a global enhancement of one polarization mode. Because the mode getting amplified depends on the redshift, we expect that an equal number of right- and left- handed modes will get amplified and attenuated, leaving the net polarization of the GW population unchanged from the vacuum scenario.
    \item Continuous, quasi-monochromatic GWs such as white dwarf binaries, or supermassive black hole binaries, which will be observed by LISA, are a promising source to search for oscillating (pseudo)scalar effects, in addition to compact binary coalescences in the ground-based regime. In this case, the intervening oscillatory field induces a time-modulation in the GW waveform. 
\end{enumerate}

In what follows, we characterize these effects for both the parity-even and parity-odd gravitational couplings. The scope of this work is to provide a theoretical description of the oscillatory features imprinted in both individual GW sources and GW populations by ultralight fields, while future work will perform more precise forecasts for applications to GW data, taking into account observational degeneracies, instrument systematics, and selection effects. If observed, these phenomena will open a new observational window to the potential zoo of ultralight scalar and pseudoscalar fields in our universe. 

\section{Parity-Even Gravitational Scalar Coupling}
\label{sec:even}
Let us first study the parity-invariant scenario in which the ultralight scalar field couples to the parity-even Gauss-Bonnet invariant. In Section~\ref{ssec:PEcorrections} we show how the GW waveform is modified. We then show that the dispersion relation is also modified and place constraints on the coupling parameter in Section~\ref{ssec:PEspeed}, and finally in  Section~\ref{ssec:PEObs} show how the oscillations of the scalar field may imprint on other GW observables.

\subsection{Gravitational-Wave Corrections: Parity Even}
\label{ssec:PEcorrections}
Let us first discuss the parity-invariant scenario, in which the ultralight scalar field, $\varphi$, couples to gravity in the following way: 
\begin{align}
S_{\rm PI} = \int d^4x \sqrt{-g}\Bigg(\kappa R &- \frac{1}{2}g^{\mu\nu}\nabla_\mu \varphi \nabla_\nu \varphi \nonumber \\
&- \frac{1}{2}m_\varphi^2 \varphi^2 + \frac{\alpha_{\pin}}{4}\varphi \mathcal{G}\Bigg), 
\label{eq:SEven}
\end{align} 
where $\alpha_\pin$ is a coupling parameter, $\kappa = (16\pi G)^{-1}$ and $\mathcal{G}$ is the four-dimensional Gauss-Bonnet invariant defined by:
\be 
\mathcal{G} = R^2 - 4R^{\mu\nu}R_{\mu\nu} + R_{\mu\nu\rho\sigma} R^{\mu\nu\rho\sigma}.
\ee 
This coupling is motivated from a variety of sources, including heterotic string theory, where the Gauss-Bonnet term arises in four dimensions upon compactification \cite{Zwiebach:1985uq, Gross:1986mw, Cano:2021rey}. It has also been shown that these parity-even quadratic curvature terms can arise from integrating out heavy particles \cite{Ema:2021fdz}.  Therefore, it is well-motivated to consider couplings of ultralight scalar fields of this form. There is a substantial body of work considering modifications of the gravitational sector when this field is massless: see the review \cite{Fernandes:2022zrq} and references therein, and a smaller subset considering implications for the massive scalar theory \cite{Nojiri:2005vv,Xie:2024xex, Chung:2026zon}. A smaller, more recent body of work has considered potential observable consequences of ultralight dark matter with such a quadratic coupling \cite{Ema:2021fdz, Dunsky:2025pvd}.

In this spirit, we can now consider the effect on gravitational waves in the situation in which the background scalar field is 1) massive and 2) oscillating, which will make the effects distinct from those in the usual formulation of scalar-Gauss-Bonnet gravity. In this case, the field can be written as: 
\be 
\varphi = \tilde{\varphi}_0(t)\cos(m_\varphi t),
\label{eq:varphi}
\ee 
where $\tilde{\varphi}_0(t)$ characterizes the amplitude of the field, and we assume that the gravitational corrections to Eq.~\eqref{eq:varphi} are sufficiently small during propagation such that they can be neglected. To determine the effect on gravitational waves, we begin by considering the gravitational-wave perturbation on a Friedmann-Lemaitre-Robertson-Walker (FLRW) metric:
\be 
ds^2 = -dt + a^2(t)(\delta_{ij} + h_{ij})d\vec{x}^2 ,
\ee 
where $a(t)$ is the scale factor, and $h_{ij}$ is the GW perturbation. On an FLRW background, $\tilde{\varphi}_0$ is given by 
\be 
\tilde{\varphi}_0(t) = \varphi_0 a(t)^{-3/2},
\ee 
where $\varphi_0$ is a constant, comoving amplitude, and the factor of $a^{-3/2}$ accounts for the dilution of the field density due to the expansion of the universe.

Upon linearizing in terms of the  gravitational-wave perturbation, $h_{ij}$ on the FLRW background and converting to conformal time, $\eta$, defined by $d\eta = dt/a$, it is known that the action, Eq.~\eqref{eq:SEven} becomes \cite{Satoh:2007gn}:  
\begin{align} 
S &= \frac{1}{8}\int d^4 x a^2\Bigg[\left(1 - \frac{\mathcal{H}\alpha_\pin\varphi'}{2 \kappa a^2}\right){h^{ij\prime}}h_{ij}'\nonumber \\
&- \left(1 - \frac{\mathcal{H}\alpha_\pin\varphi'}{2\kappa a^2} - \frac{\alpha_\pin\varphi''}{2\kappa a^2}\right)h^{ij,k}h_{ij,k} \Bigg].
\label{eq:SpertEven}
\end{align} 
Here, $'$ denotes a derivative with respect to conformal time, and $\mathcal{H}$ is the comoving Hubble parameter. From the action, Eq.~\eqref{eq:SpertEven}, we can determine the equations of motion. It is convenient to express the GW equations of motion in terms of left- and right-handed modes, $h_{\rm R,L}$. Doing so, and expanding in small $\alpha_\pin \varphi$, we find:
\be 
h''_{\rm R,L} - \left(2\mathcal{H} - \frac{\mathcal{H}\alpha_\pin \varphi''}{2\kappa a^2}\right)h'_{\rm R,L} + k^2\left(1 - \frac{\alpha_\pin \varphi''}{2\kappa a^2}\right) h_{\rm R,L}=0,
\label{eq:PIEOM}
\ee 
where we have kept only terms linear in $\varphi$, its derivatives, and $\mathcal{H}$. Directly from Eq.~\eqref{eq:PIEOM} we can appreciate that there will be both amplitude and phase corrections to the GW waveform, and that the right- and left-handed GWs will be equivalently impacted, in contrast to the next section where we will observe birefringent effects. Calculating the corrections to the GW waveform, as in \cite{Alexander:2009tp, Alexander:2017jmt, Jenks:2023pmk}, we find that indeed there are indeed contributions to both the GW phase and amplitude, determined by the coupling, $\alpha_\pin$ and scalar mass, $m_\varphi$ as follows:
\begin{widetext}
\begin{align}
h_{\rm R,L}^{\rm PI} =\bar{h}_{\rm R,L} &\times \exp\left\{ \frac{\alpha_\pin}{4}H_0 m_\varphi  \varphi_0 a(t)^{-3/2}\left[\sin \left(m_\varphi t_{\rm em}\right) - \sin \left(m_\varphi t_{0}\right)\right]\right\}\nonumber \\
&\exp\left\{-i\frac{\alpha_\pin \pi f}{2}m_\varphi \varphi_0 a(t)^{-3/2} \left[\sin \left(m_\varphi t_{\rm em}\right) - \sin \left(m_\varphi t_{0}\right)\right] \right\},
\label{eq:hRLPIt}
\end{align} 
\end{widetext}
where we have converted to physical time, $t$ and again kept only terms up to linear order in $H_0$. Here, $\bar{h}_{\rm R,L}$ corresponds to the GW modes propagating in vacuum without any scalar coupling, $t_{\rm em}$ and $t_0$ are the time of emission and observation, respectively, $f$ is the GW frequency and $H_0$ is the present-day value of the Hubble parameter. In Eq.~\eqref{eq:hRLPIt}, the amplitude correction is subdominant to the phase due to the factor of $H_0$, but for completeness, we will retain both terms in what follows. We can easily convert this into redshift, recalling that in a $\Lambda {\rm CDM}$ universe, 
\be 
t(z) = \frac{1}{H_0}\tilde{t}(z)= \frac{1}{H_0}\int_z \frac{dz}{(1+z)\sqrt{\Omega_m(1+z)^3 + \Omega_\Lambda}},
\label{eq:tz}
\ee 
where $H_0$ is the present-day value of the Hubble parameter, $\Omega_m$ and $\Omega_\Lambda$ are the energy densities of dark matter and dark energy, respectively, and we have defined $\tilde{t}(z)$ as the integral over redshift in Eq.~\eqref{eq:tz}. 

\begin{align}
h_{\rm R,L}^{\rm PI} \approx
   & \bar{h}_{\rm R,L} \exp\left\{\frac{\alpha_\pin H_0 m_\varphi  \varphi_0(1+z)^{3/2}}{4\kappa}\sin \left[\frac{m_\varphi}{H_0}\tilde{t}(z)\right] \right\}\nonumber \\
    &\times \exp\left\{-i\frac{\alpha_\pin \pi f}{2\kappa}m_\varphi  \varphi_0(1+z)^{3/2}\sin \left[\frac{m_\varphi }{H_0}\tilde{t}(z)\right]\right\}
    \label{eq:hRLPIz}
\end{align}

We can see that for a single GW event, there will be an overall amplitude correction and phase shift in the waveform, characterized by the scalar mass, $m_\varphi$, redshift, $z$, frequency, $f$, and $\varphi_0$. Notably, the signs of the corrections are redshift-dependent, indicating the overall effect on GWs will vary based on the source distance. This is significantly different than the static background scenario, in which the GW correction is directly proportional to $t(z)$ , which shifts the amplitude and phase in one direction, as a monotonic function of $z$. 

\subsection{Gravitational-Wave Speed Modification}
\label{ssec:PEspeed}
We can also appreciate that due to the phase correction, there will be a deviation in the GW speed from $v_{\rm GW} = c$. From Eq.~\eqref{eq:PIEOM}, the GW dispersion relation is given by 
\begin{align}
\omega^2 &= k^2 \left(1- \frac{\alpha_\pin\varphi''}{2\kappa a^2}\right) \nonumber \\
&= k^2\left\{1 + \frac{\alpha_\pin m_\varphi^2 \varphi_0 (1+z)^{3/2} \cos\left[\frac{m_\varphi}{H_0} \tilde{t}(z)\right]}{2\kappa}\right\}.
\end{align} 
The GW group and phase velocities are given by $v_g = d\omega/dk$ and $v_p = \omega/k$, respectively, such that  
\be 
v_g = v_p  \approx 1 + \frac{\alpha_\pin m_\varphi^2 \varphi_0 (1+z)^{3/2} \cos \left[ \frac{m_\varphi}{H_0}\tilde{t}(z)\right] }{4\kappa},\\
\label{eq:vGW}
\ee 
indicating that the GW speed itself will be impacted by the distance to the source. From Eq.\eqref{eq:vGW} it appears that the GWs can propagate superluminally. This is a common feature of effective field theories in which heavy particles are integrated out to obtain curvature-matter coupling terms, as discussed in \cite{deRham:2019ctd}. However, in many cases apparent superluminality is a feature of taking the low-energy infrared (IR) limit of a theory, while the GW speed returns to luminal in the UV, avoiding any potential issues with causality. 

\begin{figure}[htb!]
    \includegraphics[width=\linewidth]{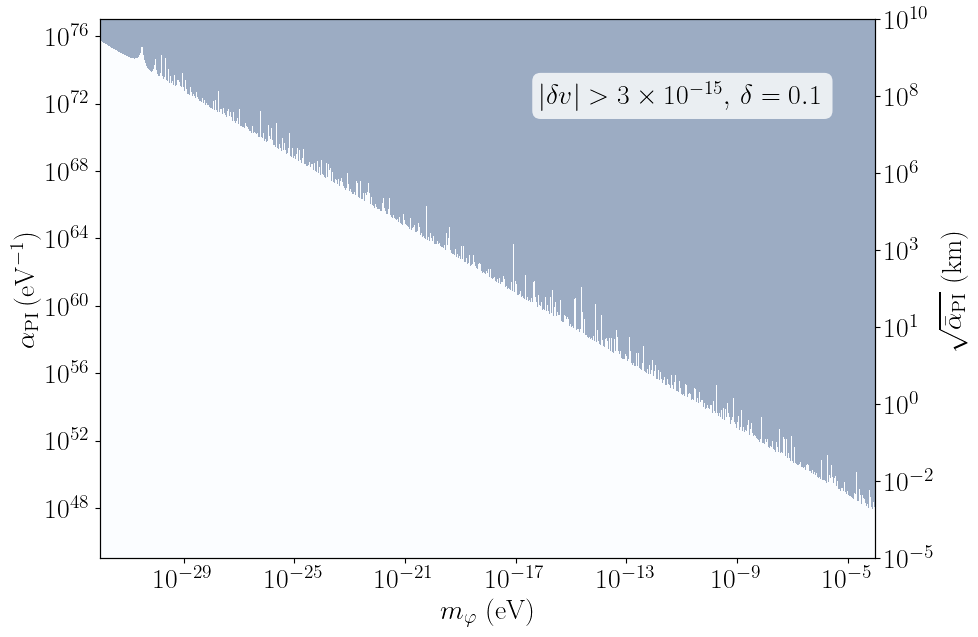}
    \includegraphics[width=\linewidth]{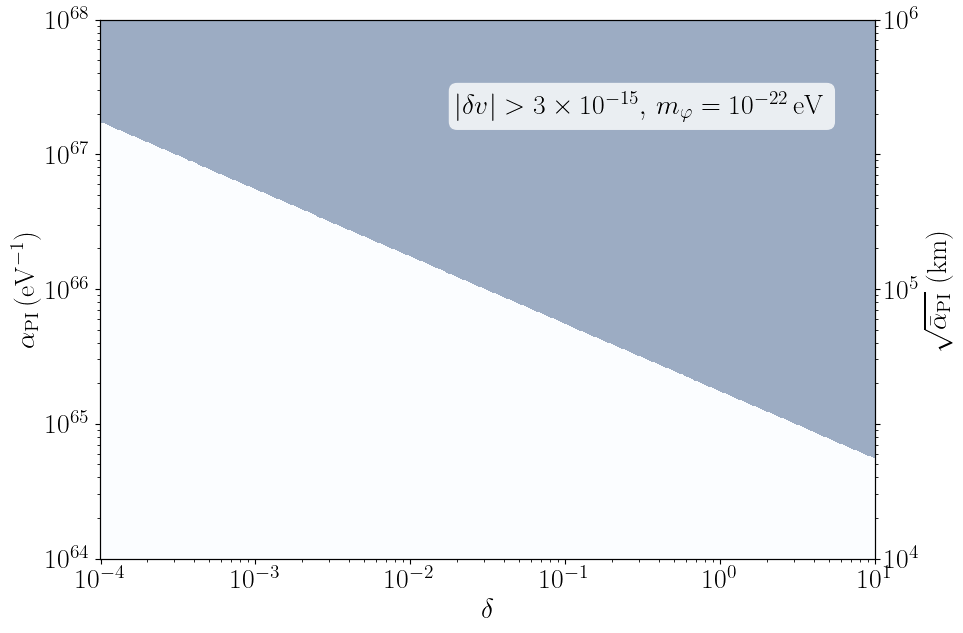}
    \caption{Joint constraint on $\alpha$ and $m_\varphi$ at fixed $\delta=0.1$ due to the observed GW speed from GW170817 (upper) and on $\alpha_\pin$ and $\delta$ at fixed $m_\varphi = 10^{-22} \, {\rm eV}$ (lower). The shaded region in each figure shows the ruled out parameter space. }
    \label{fig:vconstraint}
\end{figure}

Even if the modification in Eq.~\eqref{eq:vGW} is allowed by causality, the GW propagation speed is of course highly constrained by the coincident gravitational wave/gamma-ray burst binary neutron star event GW170817 \cite{LIGOScientific:2017ync,LIGOScientific:2017zic}. From the time delay between the GW and gamma-ray burst, the deviation of the GW speed from $c$ is constrained to be $- 7 \times 10^{-16}< |1- c_T| < 3\times 10^{-15}$, where $c_T$ is the GW propagation speed \cite{Monitor:2017mdv}. From this observation alone, we can place a joint constraint on the coupling, $\alpha_\pin$, and the mass, $m_\varphi$, depending on the oscillation amplitude, $\varphi_0$. As a benchmark, we assume that $\varphi_0$ constitutes a fraction, $\delta$, of the local dark matter density, $\rho_0 \approx 0.3 \,{\rm GeV/cm^3}$. Then, as in \cite{Fedderke:2019ajk}, this translates to 
\be 
\varphi_0 = 2.1 \times 10^{9} {\rm GeV} \delta^{1/2}\left(\frac{m_\varphi}{10^{-21} {\rm eV}}\right)^{-1}.
\ee
Taking the weaker $1- c_T$ constraint requires that
\be 
\left| \frac{\alpha_\pin m_\varphi^2 \varphi_0 (1+z)^{3/2} \cos \left[\frac{m_\varphi }{H_0}\tilde{t}(z)\right]}{4\kappa} \right| < 3 \times 10^{-15}, 
\ee 
and recalling that the GW170817 event occurred at redshift $z\approx 0.01$, we can place a constraint on the coupling $\alpha_\pin$ in terms of $m_\varphi$ and $\delta$, shown in Figure~\ref{fig:vconstraint}. The blue shaded region denotes where the GW170817 speed constraint is violated in terms of $\{\alpha_\pin, m_\varphi\}$ at fixed $\delta=0.1$ in the upper panel and in terms of $\{\alpha_\pin, \delta\}$ at fixed $m_\varphi = 10^{-22}$ eV in the lower panel. We have also shown the conversion to geometric units, $\bar{\alpha}_{\pin} = \alpha_\pin \varphi_0/\kappa$, as is customary in studies of Gauss-Bonnet gravity. The excluded region in the upper panel has jagged edges due to the oscillatory $\sin$ dependence of $v_{\rm GW}$; there are certain $m_\varphi$ values where the correction vanishes and the theory is unconstrained.  

\subsection{Gravitational-Wave Observables: Parity Even}
\label{ssec:PEObs}

\begin{figure}[htb!]
    \includegraphics[width=\linewidth]{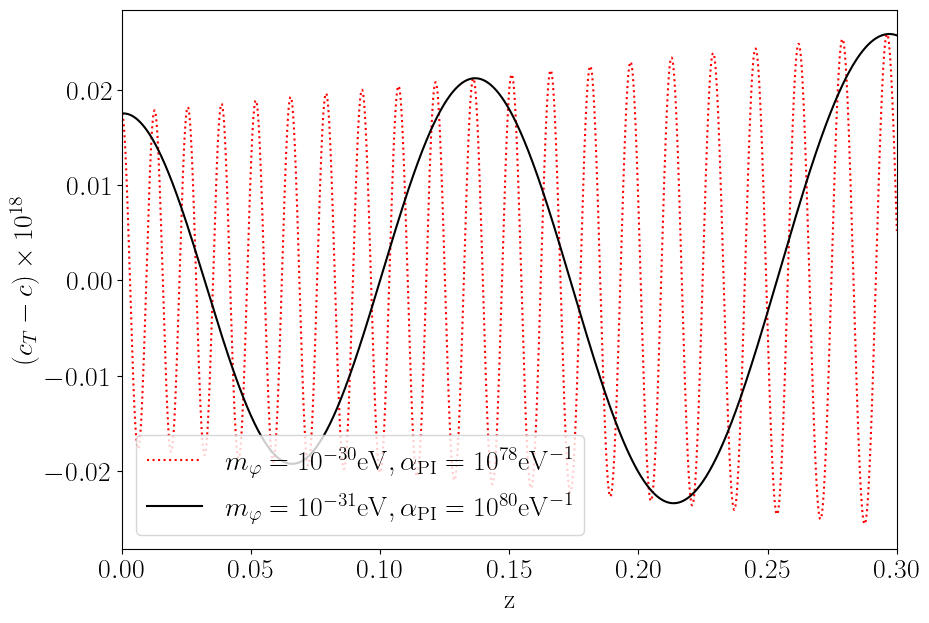}
    \caption{Oscillation in GW speed as a function of redshift.}
    \label{fig:vdist}
\end{figure}

GW170817 is a single event, but in the future with next-generation detectors we anticipate many more binary neutron star events with electromagnetic counterparts. Because the observed speed of a GW varies with the propagation distance, there will be a characteristic oscillation, with a frequency modulated by $m_\varphi$ in the speed distribution of GWs as a function of the source redshift.   Figure~\ref{fig:vdist} shows an example of this modulation for the scenario in which $m_\varphi = 10^{-31}$ eV, $\alpha_\pin = 10^{80} \, {\rm eV}^{-1}$, and $m_\varphi=10^{-30} \rm{eV}, \alpha_\pin = 10^{78} \rm{eV}^{-1}$. As a benchmark, we take $\delta \approx 0.3$ such that $\varphi$ can be subdominant fraction of the local dark matter density. Per the previous section, these values for $\alpha_\pin$ have been ruled out, but are chosen for demonstrative purposes and visual clarity in Figure~\ref{fig:vdist}. We can see that there is a clear deviation in $v_{\rm GW}$ away from $c$.

There is a clear pattern of oscillation with redshift, where the frequency is controlled by $m_\varphi/H_0$. Therefore, for heavier masses, for example near $m_\varphi \sim 10^{-22}$ eV required for $\varphi$ to account for 100\% of the dark matter, the oscillations will be too rapid to be able to resolve in redshift space. However, in this heavier mass scenario, we can still find signatures of the scalars within the GW signals. In this case, the velocity distribution will not be clearly oscillatory, but will have a statistically significant scatter away from $v_{\rm GW}/c = 1$. For example, to observe an $\approx 10 \%$ deviation from $c_T - c = 0$ at $\sim 2\sigma$, one needs to observe $\sim$ 400 events, which is within the range of proposed next-generation ground-based GW detectors such as Cosmic Explorer and Einstein Telescope. If this scatter is observed, one will then turn to analysis of individual events to determine $m_\varphi$ from the waveform distortions. We will discuss this further in the following section.  

The effects of oscillating ultralight scalars on GW propagation can also be illuminated from the GW amplitude. As can be appreciated from Eq.~\eqref{eq:hRLPIz}, the overall amplitude of the GW signal will be enhanced or attenuated based on the redshift. As with the oscillation of the speed distribution, this will imprint itself as a modulation of the `effective' redshift distribution of binary events observed by GW detectors. Unfortunately, due to the constraints on $\{\alpha_\pin, m_\varphi\}$ in Figure~\ref{fig:vconstraint} and the factor of $H_0$ in the amplitude correction, this effect is both highly constrained and suppressed to a likely unobservable level. However, for completeness and because a similar effect appears in the parity-violating case which we will discuss next, let us discuss the physical consequences anyway.    

\begin{figure}[htb!]  \includegraphics[width=\linewidth]{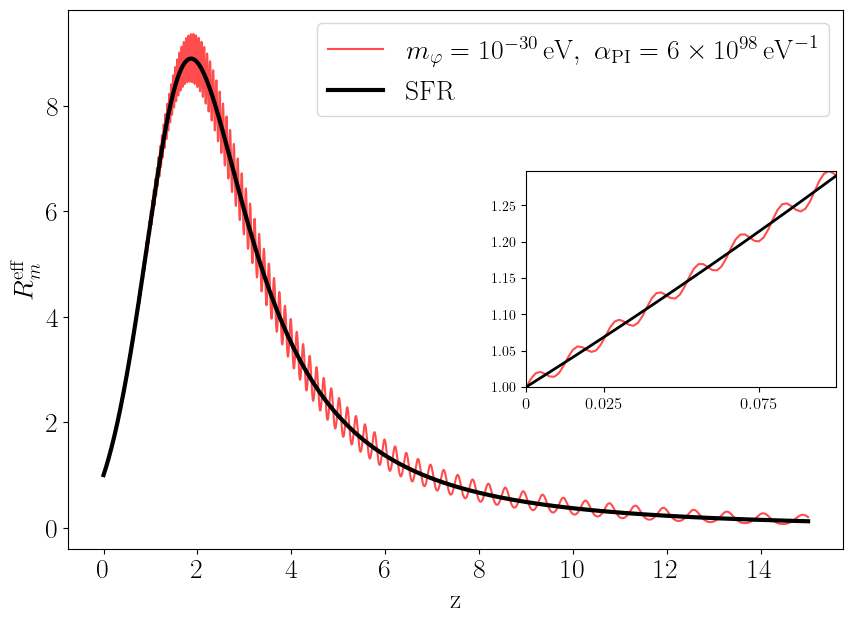}
    \caption{Modulation of the observed redshift distribution at a frequency of $m_\varphi/H_0$. The amplitude is set by the coupling parameter.}
    \label{fig:Rmmod}
\end{figure}

To highlight the physical effects, consider a simplified scenario in which the redshift distribution of GW events follows from a binary merger rate, $R_m$ which follows the global star formation rate.\footnote{This situation is highly simplified as it does not take into account any information about the sources, or GW detectors. We also expect that the BBH merger rate will follow a more complicated distribution, taking into account environmental metallicity and evolutionary time delays. However, the physical effects we want to highlight will be the same in a more realistic scenario.} Without any dark matter modulation, the star formation rate can be modeled as \cite{Madau:2014bja}:
\be 
R_m^{\rm SFR}(z) = \frac{R_0}{\mathcal{C}}
\frac{(1+z)^{\tilde{\alpha}}}{1 + \left(\frac{1 + z}{1 + z_p}\right)^{\tilde{\alpha} + \beta}},
\label{eq:RSFR}
\ee 
where $\tilde{\alpha} = 2.7, \beta=2.9, z_p = 1.9$. The normalization constant, $\mathcal{C}$ is 
\be 
\mathcal{C} = \left[1 + \frac{1}{(1 + z_p)^{\tilde{\alpha} + \beta}}\right]^{-1},
\ee 
and $R_0$ is the merger rate today. The dark matter modulated observed effective redshift distribution can then be characterized as 
\begin{equation} 
R_m^{\rm eff} \approx R_m^{\rm SFR} 
   \exp \left[ \frac{\alpha_\pin H_0 m_\varphi \varphi_0 (1+z)^{3/2}}{4\kappa}\right]\sin\left[\frac{m_\varphi}{H_0}\tilde{t}(z)\right] .
\end{equation}
A representative example of the situation is shown in Fig.~\ref{fig:Rmmod}.

Here we can see that there is an overall oscillation of the observed redshift distribution, modulated by the scalar mass, $m_\varphi/H_0$. Similar to the GW speed distribution, for heavier values of $m_\varphi$, one can search for statistically significant scatter in the observed redshift distribution to indicate the presence of an oscillating, ultralight field, then analyze individual waveforms to determine the mass and coupling. Though this is at the moment an unobservably small effect, this modulation is a clear signature of this class of gravitational couplings to an oscillating field. We will see a similar, less constrained effect in this next section.

\section{Parity-Odd Gravitational Axion Coupling}
\label{sec:odd}

Having characterized the gravitational-wave corrections to the parity-invariant dark matter coupling, let us now turn our attention to the parity-violating case. Now, we will consider a pseudoscalar gravitational axion (``gravi-axion'') field, characterized by a coupling to the parity-odd Pontryagin density. Note that this is not the QCD axion or a U(1) axion-like-particle. The gravi-axion is distinct from these other axions in that it is characterized by its parity-odd coupling to gravity.  For simplicity, we will refer to $\varphi$ as an axion or gravi-axion throughout this section. In Section~\ref{ssec:PVcorrections}, we derive the corrections to the GW waveform and in Section~\ref{ssec:PVObs} discuss the observable effects. 

\subsection{Gravitational-Wave Corrections: Parity Odd}
\label{ssec:PVcorrections}

We will consider a gravi-axion
$\varphi$ with a Chern-Simons coupling characterized by the following action:  
\be 
S_{\rm PV} = \int d^4x \sqrt{-g}\left(\kappa R - \frac{1}{2}\partial_\mu \varphi \partial^\mu \varphi - \frac{1}{2}m_\varphi^2 \varphi^2 + \frac{\alpha_\pv}{4}\varphi R \tilde{R}\right), 
\label{eq:SOdd}
\ee 
where now $\varphi$ is a \textit{pseudo}scalar field which couples via $\alpha_\pv$ to the parity-odd Pontryagin density, defined as 
\be 
R\tilde{R} = \frac{1}{2}\epsilon^{\rho\sigma\alpha\beta}R^\mu{}_{\nu\alpha\beta}R^\nu{}_{\mu\rho\sigma}.
\ee 
This type of coupling has been considered in depth when $\varphi$ is massless; see the review \cite{Alexander:2009tp} and references therein.  Recent work has also considered implications for a massive field with this type of coupling \cite{Alexander:2025olg, Ema:2021fdz, Figliolia:2025dtw}. In the massless theory, the gravitational-wave propagation effects have been precisely characterized, and lead to an overall amplitude birefringence, in that one of the right- or left-handed gravitational wave amplitudes gets enhanced, and one gets attenuated \cite{Lue:1998mq, Alexander:2007kv}. As a result, one expects a population of gravitational wave events to have an overall polarization asymmetry with a net circular polarization. In addition to the masslessness of the field, this effect relies on the assumption that the background field is slowly varying and effectively a constant, such that one can approximate $\dot{\varphi} \approx \dot{\varphi}_0$, where ${\varphi}_0$ is the present-day value of the field and $\dot{\varphi}$ is the derivative of the field in physical time. As in the parity-event coupling scenario discussed previously, we will now extend this analysis to the scenario in which the background axion field is massive and oscillating. 

From the action, Eq.~\eqref{eq:SOdd}, we can again characterize the tensor perturbation, $h_{ij}$ on a FLRW spacetime to find the modified linearized equations of motion \cite{Alexander:2009tp}:
\be 
    \Box h^j{}_i = -\frac{\alpha_\pv}{\kappa a^2}\epsilon^{pjk}\left[(\varphi'' - 2\mathcal{H}\varphi')\partial_p h_{ki}' + \varphi' \partial_p\Box h_{ki}\right].
\ee 
Introducing a plane-wave ansatz:
\be 
h_{ij} = A_{ij} e^{-i(\phi(\eta) - \Vec{k}\cdot\Vec{x})},
\ee 
we then obtain the dispersion relation for $\phi$, which can be linearized as 

\be 
\phi(\eta) = \bar{\phi}(\eta) + \delta\phi(\eta).
\ee 
Converting to physical time, $t$, we obtain
\be  
\delta\phi = \frac{-i\alpha_\pv\lambda_{R,L}f\pi}{\kappa} \int \left( \ddot{\varphi} - H\dot{\varphi}\right)dt ,
\ee   
for the axion-induced perturbation. 
With the axion field profile as in Eq.~\eqref{eq:varphi}, and assuming that $m_\varphi \gg H_0$, we have 
\begin{align} 
h_{\rm R,L}^{\rm PV} &\approx \bar{h}_{\rm R,L}\nonumber \\
&\times \exp\left[\pm \frac{\alpha_\pv f\pi\varphi_0 a(t)^{-3/2}}{\kappa} m_\varphi \left(\sin m_\varphi t_{em} - \sin m_\varphi t_0\right) \right].
\label{eq:hPVt}
\end{align} 

As in the previous section, we can equivalently write this as 
\begin{align} 
h_{\rm R,L}^{\rm PV} \approx \bar{h}_{\rm R,L}\exp\left\{ \pm \frac{\alpha_\pv f\pi \varphi_0(1+z)^{3/2}}{\kappa} m_\varphi \sin\left[\frac{m_\varphi}{H_0}\tilde{t}(z)\right]\right\}
\label{eq:hPVz}
\end{align}
for a $\Lambda{\rm CDM}$ universe, where $\tilde{t}(z)$ is again defined by Eq.~\eqref{eq:tz}.

There are several differences here compared to the parity-invariant scenario discussed in Section~\ref{sec:even}. First, the GW phase remains unchanged and we only have an amplitude correction. We can also explicitly see the birefringence between the right- and left-handed modes that is not present for the parity-even scalar coupling, which arises due to the parity-odd nature of the pseudoscalar axion. This birefringence happens as well in the constant-background $\varphi$ scenario, however we now have the additional characteristic oscillation in redshift. With a constant, massless axion background, all of the right- or left-handed GWs get amplified or attenuated. In the massive and oscillating axion case, the amplified mode depends on the emission redshift. As a result, we expect a similar polarization `washout' effect that has been found for the CMB \cite{Fedderke:2019ajk}, if we look at the net population-level polarization. Because the sources are randomly distributed, we no longer expect an overall enhancement of one polarization in the population. This is distinct from the constant background scenario, in which one expects there to be a global overall preference for right- or left-handed circular polarization. 

\subsection{Gravitational-Wave Observables: Parity Odd}
\label{ssec:PVObs}

\begin{figure}
    \includegraphics[width=\linewidth]{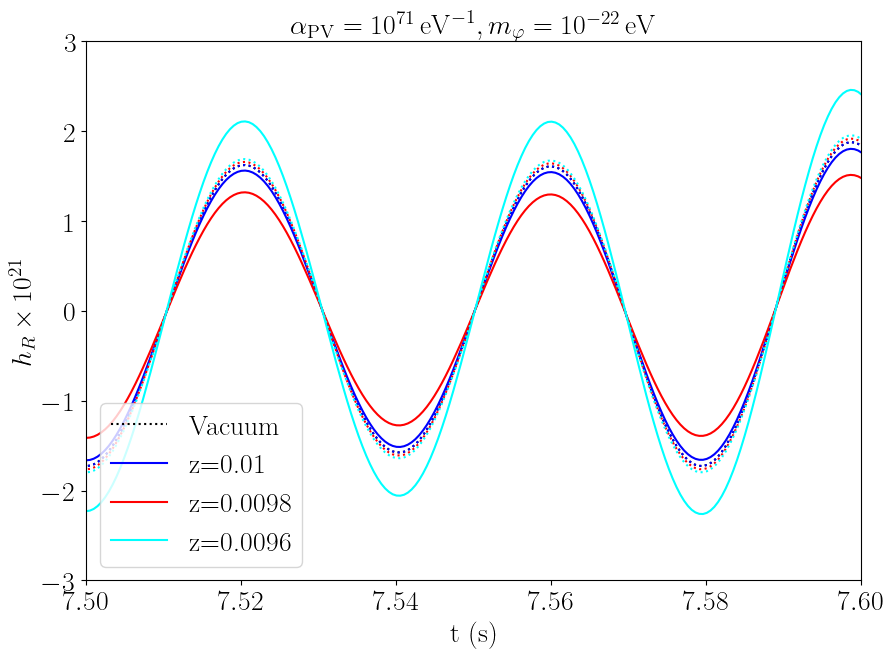}
    \includegraphics[width=\linewidth]{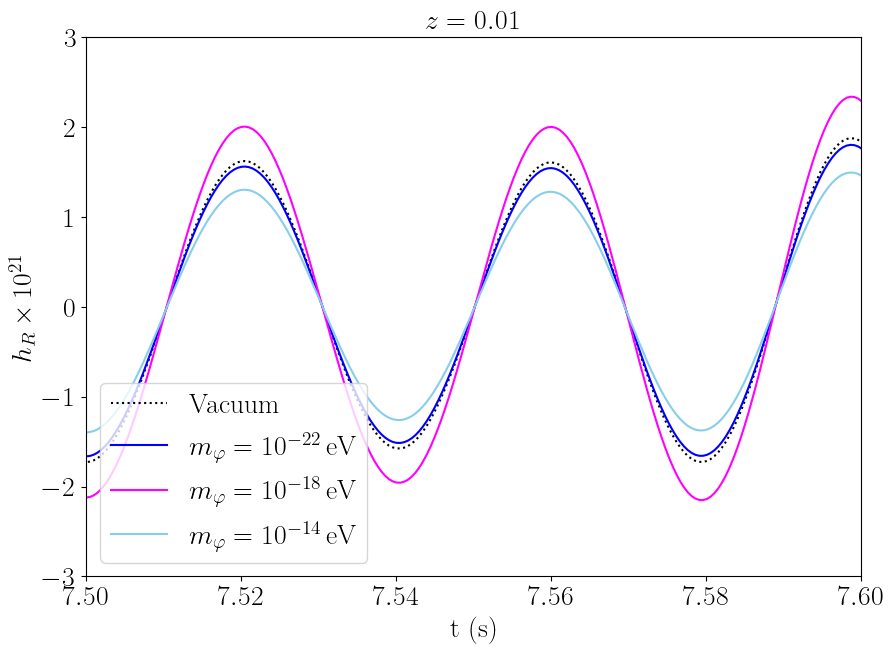}
    \caption{Parity-violating waveform modification for GWs at fixed mass and varying redshift (top) and fixed redshift with varying mass (bottom). The dotted lines correspond to propagation in vaccuum with no coupling to an ultralight field/ In the upper panel, we show the modification at fixed $m_\varphi = 10^{-22}$ eV at three different redshifts, and the lower panel shows the effect of varying $m_\varphi$ at fixed redshift, keeping $m_\varphi \alpha_\pv= 10^{49}$ constant across all three cases.  }
    \label{fig:hRmod}
\end{figure}

Let us now investigate the effect of the waveform modifications in Eq.~\eqref{eq:hPVz} on observables. First,
to demonstrate the effects on individual events, Figure~\ref{fig:hRmod} shows a single waveform simulated using PyCBC \cite{pycbc2016} for an equal mass binary black hole system with various axion modifications. The system has $m_1 = m_2 = 30 M_{\odot}$ and is chosen to have an inclination of $\iota = \pi/2$ such that $h_R = h_L$ when propagating in vacuum. 

In the top panel, we show the difference in amplitude for a GW interaction with an axion with $m_\varphi = 10^{-22}$ eV. We consider three identical events, at very close redshift such that the unmodified waveforms are nearly identical. Depending on the source redshift, the modified waveform is either amplified or attenuated. At $z=0.0096$, the GW amplitude is amplified, but at $z = 0.0098$ and $z= 0.01$ it is attenuated compared to the free propagation scenario. In the bottom panel, we show the effects at fixed redshift for varying $m_\varphi$, set such that $\alpha_\pv m_\varphi$ is constant across all three cases. Similarly, the overall amplification/attenuation of the GW is not constant as one varies the mass. Therefore, a single event with no additional electromagnetic information provides a challenge for extracting information about the axion mass and coupling. However, for enough events with precisely measured redshifts, one can extract the associated scalar mass from this waveform modulation by analyzing many individual waveform distortions. Note that for the example we have shown, the right- and left- handed GW polarizations are equivalent in the unmodified scenario, and in each of the modified cases, $h_L$ will be affected in the opposite direction to what is shown for $h_R$.

\begin{figure}
    \centering 
    \includegraphics[width=\linewidth]{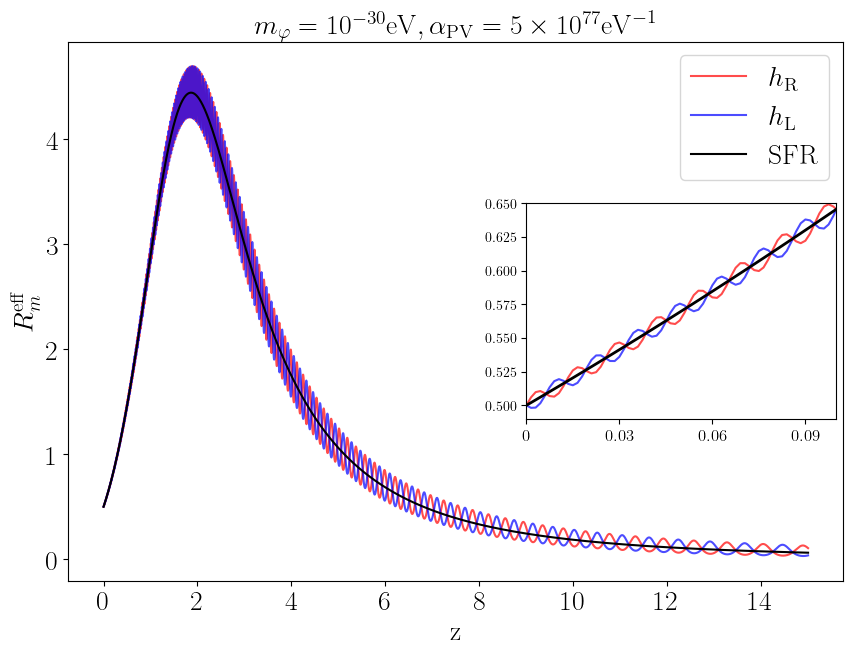}
    \caption{Parity violating oscillation effect in the observed GW population, approximated by the BBH merger rate. We assume that the vacuum distribution has equivalent contributions in right- and left-handed GWs.}
    \label{fig:RmPV}
\end{figure}

We also expect there to be a modulation in the redshift distribution of various observables. As in the parity-invariant case, we can also characterize the amplitude modulation in terms of the redshift distribution. In this scenario, the effects will be more subtle than in the parity-invariant case. Rather than in the redshift distribution as a whole, we will now have a modulation in the distribution of the right- and left- handed polarizations of the population. With polarization reconstruction of right- and left- handed amplitudes, these effects would be evident from the full GW population. Once again using the SFR, Eq.~\eqref{eq:RSFR} as a simple proxy for the overall expected observed GW amplitude, we show this explicitly in Figure~\eqref{fig:RmPV}.

We can see that there is again a clear oscillatory feature in redshift, with an offset between the right- and left- handed modes; where $h_R$ is amplified, $h_L$ is suppressed and vice-versa. Again, the oscillations will not be resolvable for heavier masses, but one can use a joint analysis with the scatter in the redshift distribution and individual waveforms. As before, this is a simplistic toy model for the observable, and also depends on being able to precisely construct the polarization amplitudes. However, this characteristic oscillation is a clear and unique signature for GW interactions with an oscillating field. 

We furthermore expect to be able to observe this effect in the redshift distribution of inclination angles and luminosity distances of GW sources. In general, we expect the inclination of binaries to be randomly oriented (up to selection effects). In our axion scenario, given that the observed inclination and luminosity distance are degenerate with the amplitude, they will also exhibit the characteristic modulation. This is an observation where multi-messenger GW events will be helpful, given that electromagnetic information can be used to precisely measure the inclination and distance of the binary in order to reduce uncertainties and compare to the observed GW properties. 

For example, consider the parameter $\xi = \cos\iota$ where $\iota$ is the inclination angle of the binary. Because the effect of the intervening axion is to alter the polarization amplitudes, this also corresponds to a change in the \textit{observed} inclination angle, assuming that the waveforms are not accounting for the axion coupling. In this case, the effective, observed $\xi_{\rm eff}$ will be \cite{Jenks:2023pmk}
\be 
\xi_{\rm eff} = \bar{\xi} + \frac{1}{2}(1 - \bar{\xi}^2)\frac{\alpha_\pv f \pi \varphi_0(1+z)^{-3/2} m_\varphi}{\kappa}\sin\left[\frac{m_\varphi}{H_0}\tilde{t}(z)\right]
\ee 
where $\bar{\xi}$ is the intrinsic value of $\xi$ before accounting for polarization birefringence. This can be observed with multimessenger effects, in which the true value of $\xi$ is measured from electromagnetic information, and compared with $\xi_{\rm eff}$ measured at GW detectors, as discussed in \cite{Yunes:2010yf, Lagos:2024boe}. Then, as a result of the $\sin$ dependence in the modification, there will be an overall oscillatory feature in $\xi_{\rm eff}$

\begin{figure}[htb!]
    \includegraphics[width=\linewidth]{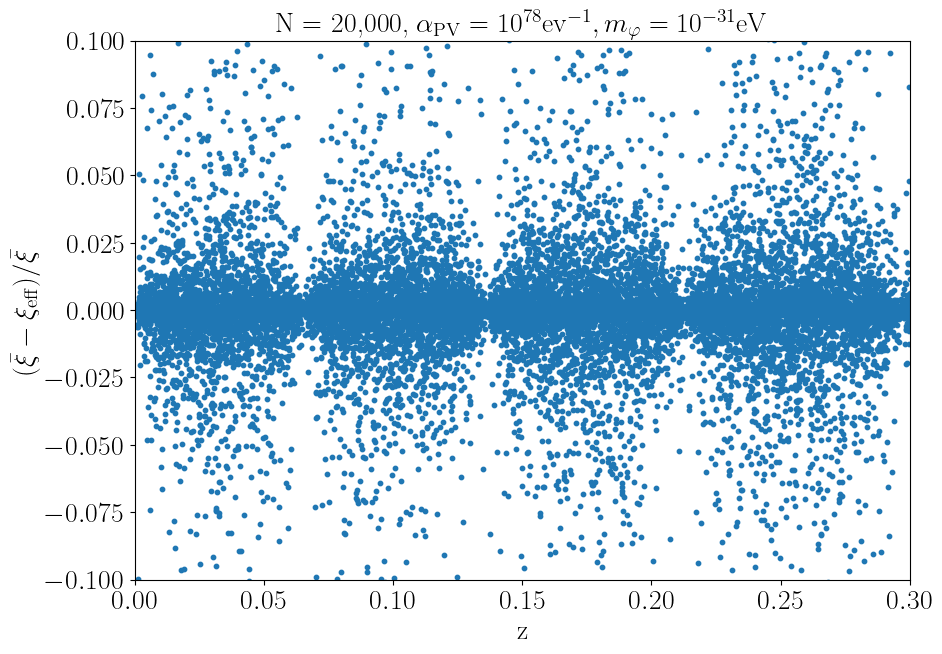}
    \caption{Fractional difference in the inclination $\iota$ from a randomized distribution as a function of redshift.}
    \label{fig:iotadist}
\end{figure}

Figure~\ref{fig:iotadist} shows an example of this observed inclination distribution, characterizing the fractional deviation in $\xi_{\rm eff}$ compared to a set of $\bar{\xi}$ chosen such that the inclination angles are randomly distributed from $-\pi < \iota < \pi$. 

For demonstrative purposes we show $20,000$ events, but it will be possible to obtain statistically significant deviations from a random scatter with far fewer events, on the order of several hundred, which may be feasible with next-generation detectors such as Cosmic Explorer and Einstein Telescope. Furthermore, it will be important to carefully take detector systematics and selection effects into account when performing any analysis on observations of $\xi$.
 
As in the parity-even scenario, observing population-level oscillations in various parameter distributions requires very low axion masses, such that the characteristic oscillations can be resolved in redshift space, while a heavier field will induce a statistically significant scatter in population observables. Therefore, as in the parity-even scenario, in the regime where the oscillations are too rapid to resolve, one can still extract information about $m_\varphi$ by looking for scatter in the observed amplitude, inclination, or luminosity distance distributions to indicate the presence of an oscillating field, then analyze individual waveforms to determine the precise mass and coupling.

\section{Time Modulation of Continuous Sources}
\label{sec:monoGW}
In the above two sections, we considered compact binaries in the ground-based detector regime. However, the modulation by ultralight fields will be present in other scenarios as well. In particular, LISA will be well-situated to observe these effects. Expected to launch in 2035, LISA will be sensitive GW signals in the mHz frequency range and will observe a wide range of signals, including GWs from galactic binaries, merging intermediate mass and massive black hole binaries, extreme mass ratio inspirals, and possibly even cosmological GW signals from the early universe. If there is an ultralight field with a quadratic coupling to gravity, GWs from all of these sources will also experience a modulation from the field oscillation. In particular, some of the LISA sources, including galactic binaries and supermassive black holes in their early inspiral, will emit continuous, quasi-monochromatic GWs, which will provide an ideal laboratory for probing gravitationally coupled ultralight fields. In particular, we find that a continuously observed GW will undergo a modulation in the time domain, corresponding to the (pseudo)scalar mass. This is in contrast to the previous two sections, where the dominant observational consequence is in the observed redshift distributions of GW source observables.

To model this effect, lets consider a continuous, monochromatic source. The  polarization modes for such systems can be analytically approximated as \cite{Cornish:2005hd}: 

\begin{align}
\bar{h}_+(t) &= A_0 (1 + \cos^2\iota)\cos(\Phi(t) - \Phi_0),\\
\bar{h}_\times (t) &= - 2 A_0 \cos\iota \sin(\Phi(t) - \Phi_0), 
\end{align}
where $A_0$ is a constant amplitude, $\Phi_0$ the initial phase, and $\Phi(t)$ is the GW phase, defined in terms of the quasi-monochromatic frequency, $f_0$, its time derivative, and the position of the binary. First, consider the parity-odd case. We can apply the axion correction as in Eq.~\eqref{eq:hPVt} to obtain the corrected waveform for the monochromatic GWs, $h^{\rm mono}_{\rm R,L}$:
\begin{widetext}
\begin{align}
    h_{\rm R,L}^{\rm mono} &= \frac{A}{2\sqrt{2}} \left\{(1 + \cos^2\iota) \cos[\Phi(t) - \Phi_0] 
    \pm 2 i \cos\iota \sin [\Phi(t) - \Phi_0]\right\}\nonumber\\
    &\times \exp\left(\pm \frac{\alpha_\pv f \pi \varphi_0a(t)^{-3/2} m_\varphi}{\kappa} \left\{ \sin[m_\varphi t_{em}] - \sin[m_\varphi(t_em + D/c)]\right\}\right),
    \label{eq:MGW}
\end{align}
\end{widetext}
where $D$ is the distance to the source, $c$ is the speed of light and we have converted from the $+,\times$ basis to ${\rm R,L}$ via
\begin{align}
    h_+ &= \frac{h_{\rm R} + h_{\rm L}}{\sqrt{2}}\\
    h_\times &= i \frac{h_{\rm R} - h_{\rm L}}{\sqrt{2}}
\end{align}

In this case, we will only care about the modulation in the time domain, rather than converting to redshift. Let's consider galactic binaries an example scenario. These binaries are known quasi-monochromatic sources that are abundant in our galaxy and primarily consist of white dwarfs, though some also contain black holes or neutron stars. LISA is expected to detect and resolve thousands of these binaries during its observing run \cite{LISA:2017pwj}, some of which will also be observed electromagnetically, e.g. \cite{2018MNRAS.480..302K}. Additionally, because these sources all lie within the galaxy, any propagation distance traversed is well within the coherence length of ultralight field in the mass range we are considering. Therefore, galactic binaries can be used to search for GW modulation by an ultralight field. 

\begin{figure}[htb!]
\includegraphics[width=\linewidth]{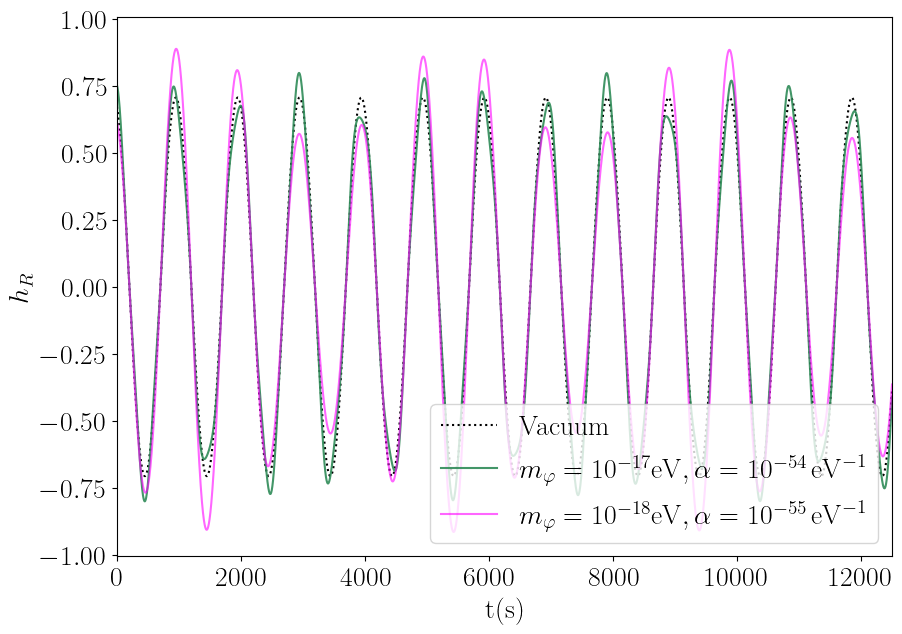} 
\includegraphics[width=\linewidth]{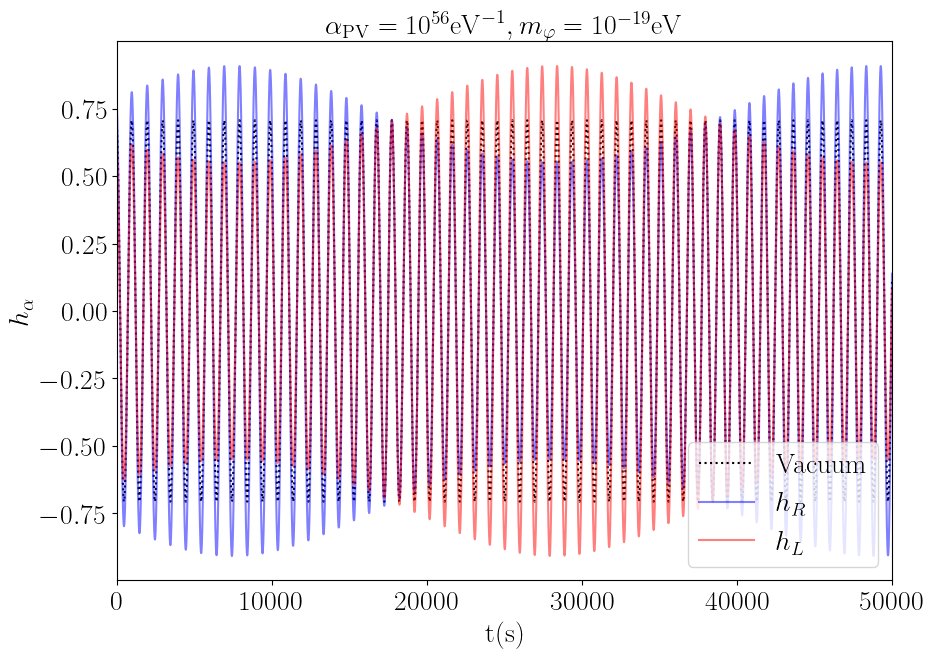}
\caption{Waveform modulation in the right-handed polarization for an axion with masses $10^{-17}, 10^{-18} {\rm eV}$ (top) and modulation of both the right- and left-handed modes for $m_\varphi = 10^{-19}$ eV (bottom).}
\label{fig:tmod-PV}
\end{figure}

 Figure~\ref{fig:tmod-PV} shows an example of the physical effects induced by the oscillating axion for three masses in the ULDM range. We consider an equal mass white dwarf binary with $m_1 = m_2 = 0.6 M_\odot$, at a distance of $d \sim 1.8$ pc, inclined at $\pi/2$ such that the right- and left-handed polarization amplitudes are the same in vacuum. In the top figure we compare the waveform right-handed polarization mode in vacuum with modulation by axions with $m_\varphi = 10^{-17}$ eV (green) and $m_\varphi = 10^{-18}$ eV (pink). In the bottom figure we show the differing modulation of the right- and left-handed modes for an intervening axion with $m_\varphi = 10^{-19}$ eV. In this case, it is clear that the right- and left- handed modes experience both an enhancement and a suppression throughout their propagation compared to the vacuum scenario, but at any given time the two polarization modes will be experiencing the opposite effect. In all three cases, the overall modulation of the frequency is controlled by the the axion mass, and the lighter the axion, the longer the period of modulation is. The timescale for such effects is such that the modulation for a wide range of masses can easily be probed with an instrument like LISA, which aims to have a four-year observing run \cite{Reitze:2019iox}. 

\begin{figure}[htb!] 
\includegraphics[width=\linewidth]{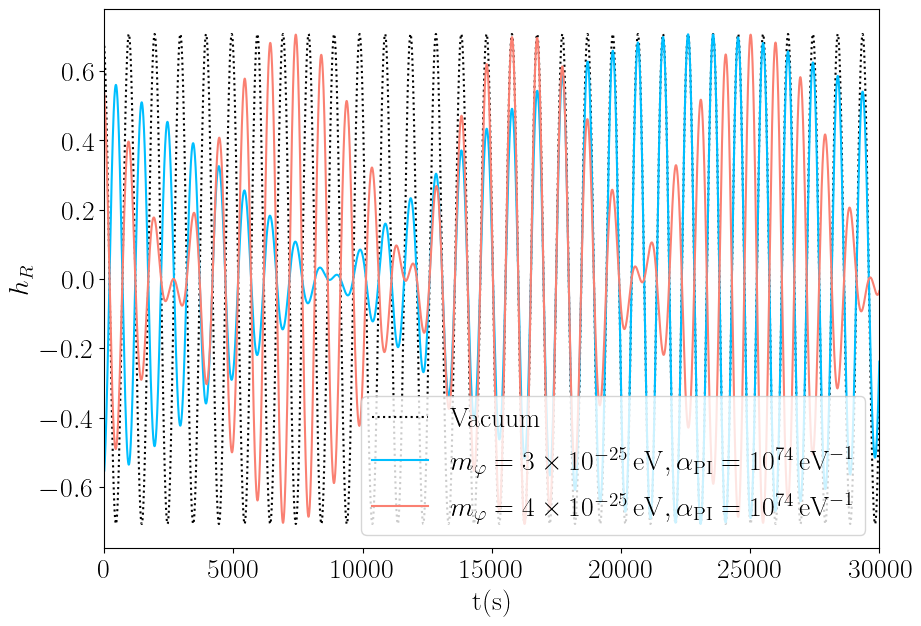}
    \caption{Phase modulation of a continuous wave by an ultralight scalar with $m_\varphi = 10^{-25}$ eV and $m_\varphi = 3\times 10^{-25}$ eV.}
    \label{fig:mono-phaseGB}
\end{figure}

As discussed in Section~\ref{sec:even}, the parity-even Gauss-Bonnet coupling is highly constrained by observations. However, for completeness, let us also characterize how a continuous monochromatic wave is modulated by an intervening ultralight scalar as a result of the parity-even coupling. Figure~\ref{fig:mono-phaseGB} shows the same binary white dwarf, quasi-monochromatic, continuous wave described above. We compare the effects of a scalar with $m_\varphi = 3\times 10^{-25}$ eV and $m_\varphi = 4\times 10^{-24}$ eV, and we can see that the dominant effect is a phase modulation controlled by $m_\varphi$, as well as a subdominant amplitude correction. Here, because there is no birefringence, the effects on $h_{\rm R}$ and $h_{\rm L}$ will be equivalent. 

From Figures~\ref{fig:tmod-PV} and ~\ref{fig:mono-phaseGB}, we can appreciate that ultralight fields with these gravitational couplings leave a striking imprint in monochromatic, continuous GW signals. While we have specifically considered a white dwarf binary system, this is not the only source where such effects will arise. We also expect to see this modulation for other known monochromatic sources such as binary supermassive black holes. Furthermore, in the above we have not taken into account a full characterization of the detector response, or the analysis procedures required to isolate one continuous GW signal and disentangle the (pseudo)scalar modulation from other effects. We defer these studies to future work. 

\section{Discussion and Conclusions}
\label{sec:discussion}

In this paper we have studied the effects of coherently oscillating ultralight fields that couple to gravity on the propagation of gravitational waves. We have shown that, in contrast to the massless, slowly varying field scenario, gravitational waves propagating through such a medium inherit a mass-dependent modulation which imprints on individual event waveforms and in the redshift distribution of various population observables. We have considered this effect for the parity-even Gauss-Bonnet coupling to a scalar and the parity-odd Chern-Simons coupling to a gravi-axion. In the parity even case, the corrections are highly constrained by the GW170817 binary neutron star event. However, we showed that GW observables such as the observed amplitude and speed inherit characteristic oscillatory features in the redshift distribution, due to the oscillating scalar field. In the parity-odd scenario, the effects are birefringent and appear in observables including the redshift distribution of the amplitude of the right- and left-handed polarization modes, as well as the observed inclination distribution of events. For both parity-even and parity-odd scenarios, we discussed that the oscillations can be observed directly in redshift space, for masses $10^{-29} \lesssim m_\varphi/{\rm eV} \lesssim 10^{-31}$. On the other hand, fields with heavier masses whose induced oscillations cannot be resolved will lead to a statistically significant scatter in the redshift distribution of observables, which can then be further probed by analyzing individual events. Finally, we have discussed continuous, monochromatic GWs as a new probe of these effects. In particular, a continuous GW will undergo modulation in the time domain, at a frequency controlled by the (pseudo)scalar mass. We modeled this for a binary white dwarf system, which are abundant in our universe and expected to be observed by LISA. These oscillatory effects of coherently oscillating fields coupled to gravity on gravitational waves provide a novel probe of ultralight fields that may live in the dark sector.

This work has primarily focused on the conceptual aspects of the effect of gravitationally coupled ultralight fields on GWs, showing how the GW waveforms and population observables are modified. These effects are straightforward to apply to existing LIGO-Virgo-KAGRA data for both individual events and the current population, as has been done for the slowly varying massless scalar \cite{ Ng:2023jjt, Callister:2023tws, Lagos:2024boe}. A full forecast for next-generation ground based detectors will be useful to determine the prospects for measuring or constraining these effects more precisely and further analysis is required to understand how to extract the full information about the modulation of a single continuous wave source from future LISA data. In future work, we will carry out these analyses and others.

\acknowledgements
LJ is supported by the Provost's Postdoctoral Fellowship at Johns Hopkins University. This work was supported by NSF Grant No.\ 2412361, NASA ATP Grant No.\ 80NSSC24K1226, and the Templeton Foundation. 

\bibliography{master}

\end{document}